\documentclass[runningheads]{llncs}
\usepackage[T1]{fontenc}
\usepackage{graphicx}
\usepackage[usenames,dvipsnames,svgnames,table]{xcolor}
\usepackage{tikz,pgfplots,pgfplotstable,amsmath}
\usepackage{multirow}
\usepackage{array}
\usepackage{tabularx}
\usepackage{tabu}                      
\usepackage{booktabs}                  
\usepackage{lipsum}                    
\usepackage{mwe}    

\usepackage{mathptmx}                  
\definecolor{color1}{RGB}{152,153,250}
\definecolor{color2}{RGB}{177,178,250}
\definecolor{color3}{RGB}{250,205,202}
\definecolor{color4}{RGB}{250,234,160}
\definecolor{color5}{RGB}{165,250,197}
\definecolor{NavyBlue}{RGB}{0,0,128}
\usepackage{xcolor}
\usepackage{todonotes}
\usepackage{graphicx}
\usepackage{csquotes}
\usepackage{soul}

\newcommand{\proj}{\textbf{SnapNCode}}

\begin{document}
\title{SnapNCode: An Integrated Development Environment for Programming Physical Objects Interactions}
\titlerunning{SnapNCode}
%
\author{Xiaoyan Wei\orcidID{0000-0001-5535-4197} \and
Zijian Yue\orcidID{0009-0009-2538-5836}
\and
Hsiang-Ting Chen\orcidID{0000-0003-0873-2698}}
\authorrunning{Wei, Yue et al.}
%
\institute{The University of Adelaide, Adelaide 5007, AU \\
\email{\{xiaoyan.wei, tim.chen\}@adelaide.edu.au\\
\{zijian.yue\}@student.adelaide.edu.au}}
\maketitle              

\begin{figure}
  \centering
  \includegraphics[width=\linewidth]{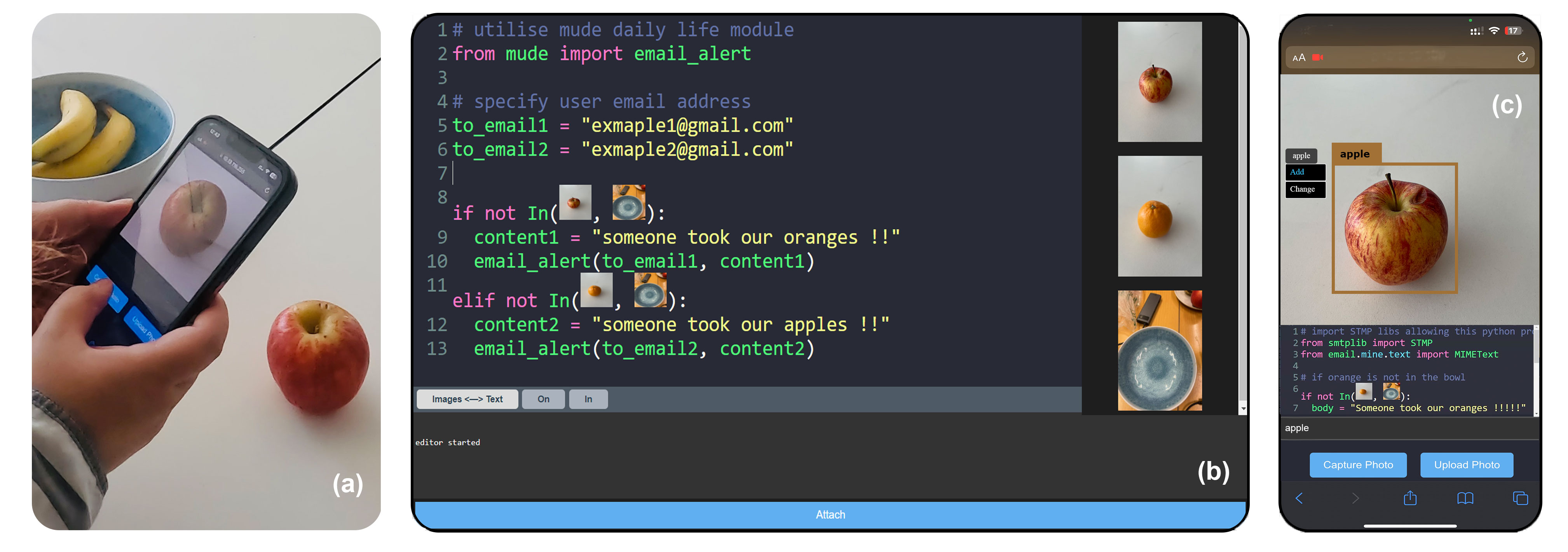}
  \caption{\proj{} is an IDE facilitating the development of spatial computing applications. (a) captures the images of an apple and send capture object to the system by using \proj{} mobile application (b) developing a notification program advising users after fruit was removed, by \proj{} pc application (c) checking existed code attached to an apple by using \proj{} mobile application.}
  \label{fig:teaser}
\end{figure}

\begin{abstract}
Spatial computing technologies have the potential to revolutionize how we interact with the world around us. 
However, most modern integrated development environments (IDEs) have not fully adapted to this paradigm shift. 
For example, physical 3D objects in the real world are still represented as 2D text variables in code, creating a significant perceptual distance between these representations. 
In response to this challenge, we introduce \proj{}, a novel IDE for spatial programming. 
\proj{} enables programmers to capture various states of physical objects through live video streams from cameras and directly insert these visual representations into their code. 
Moreover, users can augment physical objects by attaching code snippets onto objects, which are opportunistically triggered when observed by cameras. 
We conducted a user study (N=12) to assess the usability of \proj{}. 
Feedback from participants indicates that the system is easy-to-use and holds promise for daily casual uses and integration into a broader range of workflows.

\end{abstract}
\section{Introduction}
Spatial computing is an emerging research domain focused on bridging digital technologies with the physical environment, allowing computers to perceive and operate within 3D spaces. The field has gained momentum through recent advancements in machine learning (ML) and the advent of cost-effective head-mounted displays featuring high-quality video passthrough technology. Now marks a transformative phase for spatial computing, promising to boost productivity in industries like media, architecture, and manufacturing. 

As the shift towards the spatial computing paradigm continues, there is an increasing need for new content, driving the research and development of innovative content authoring methodologies.
For examples, Monteiro et al. \cite{monteiro2023teachable} introduced an novel augmented reality (AR) tool that uses vision-based interactive machine teaching to allow users to create interactive, tangible AR prototypes with everyday objects without programming, overcoming the limitations of marker-based methods.
Zhu et al. \cite{Zhu2023-IoTVR} proposed a VR learning environment that helps students gain IoT knowledge by designing, programming, and exploring real-world IoT scenarios, like a smart house, using a custom 3D block-based visual programming tool in an immersive environment.
These developments represent significant advances in authoring environments for spatial applications. 
However, most programmers continue to utilize current generation IDE, which are mostly text-based for 2D displays, for developing and deploying spatial applications due to existing workflow integration challenges. The innovative features of these research prototypes may not be easily adapted into existing IDEs.

\begin{figure} [h]
    \centering
    \includegraphics[width=\columnwidth]{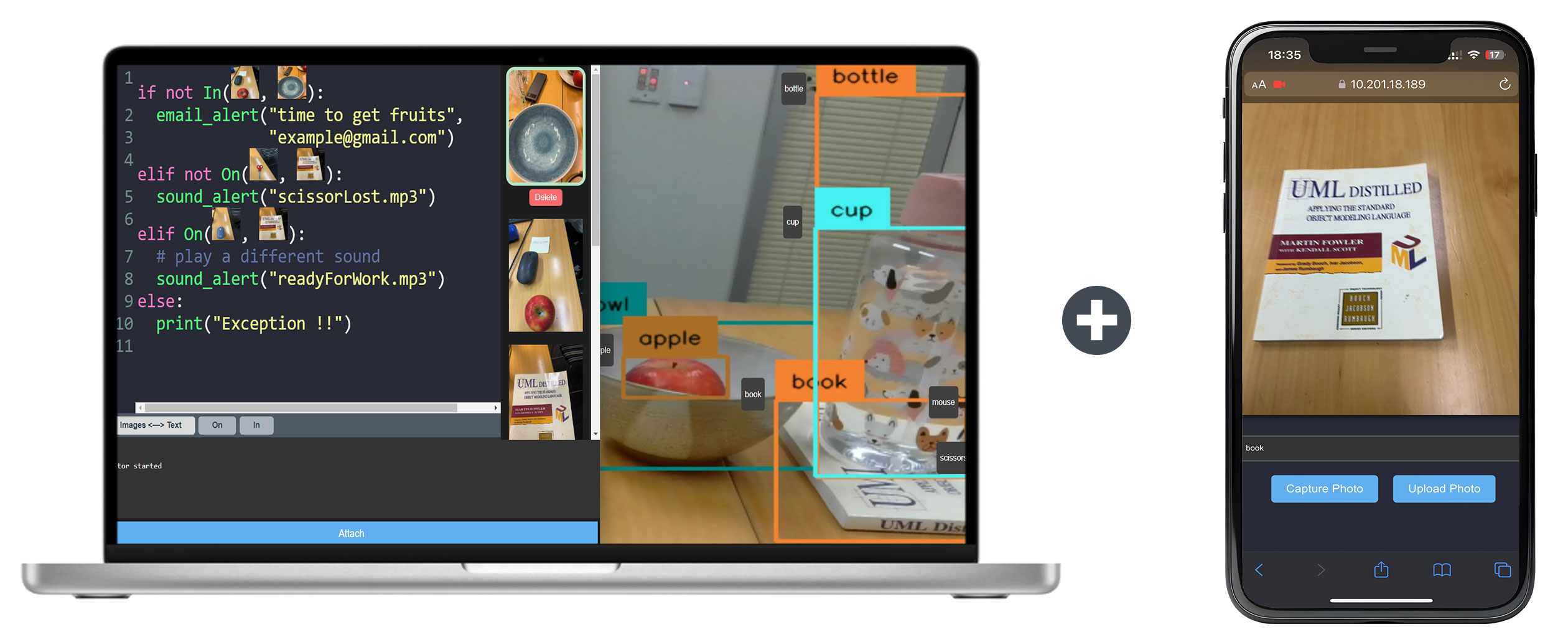}
    \caption{User Interface}
    \label{figure:interface}
\end{figure}

We propose \proj{}, a prototype coding environment designed to facilitate the development of spatial computing applications that often involves physical objects. 
Inspired by the seminal work Sikuli~\cite{Yeh2009-Sikuli}, the core idea is to use a computer vision approach to recognize and capture the states of physical objects in the environment, then allow the programmer to directly \textit{insert} these states into the code as images (see Figure \ref{figure:interface}). 
\proj{} enables users to view variables, which represent physical objects, as images, while maintaining a text-based underlying code structure to ensure compatibility with existing workflows. 
Furthermore, \proj{} facilitates the creation of event-driven code by allowing users to \textit{attach} code snippets to physical objects. 
The attached code would be triggered by changes in the object’s state, opportunistically detected either through a mobile phone camera or cameras integrated into VR/AR headsets.

We conducted a user study with 12 programmers, organized into 6 pairs, to evaluate \proj{}. 
Participants were assessed through video recordings, post-hoc questionnaires, and semi-structured interviews. 
The result suggests that participants found \proj{} simplifies the coding process involving physical objects by offering a more intuitive representation to specify the relationships between these objects. Additionally, participants agreed that the interaction introduced in \proj{} could be integrated into their existing workflows without significant modifications. 
They also noted the potential applicability of our system across various fields, including medicine, chemistry, and architecture, in the future.

The contributions of this paper are as follows:
\begin{itemize}
    \item We introduce a novel IDE prototype that allows the representation of variables corresponding to physical object states as images, while preserving the underlying text structure for easy integration into existing workflow.
    \item We propose a new concept of physical object-oriented programming that enables code to be attached to physical objects and triggered opportunistically by cameras on mobile devices or wearables.
    \item Our user study provides valuable insights into the features deemed necessary and beneficial for the development of next-generation IDEs tailored for spatial computing.
\end{itemize}
\section{Related Work}


Recent advancements in spatial computing have explored the integration of programming activities into three-dimensional environments, with the aim of strengthening the connection between the physical world and virtual environments, thus offer~\cite{Bau2017-visualprogramming}.
For example, 
Ivy~\cite{Ens2017ivy} provides a virtual reality programming tool that simplifies programming and debugging of smart objects by connecting them and visualizing real-time data flows.
FlowMatic~\cite{zhang2020flowmatic} further enhance the development enviornment by enabling programmers to create reactive behaviours and manage programming primitives directly in the virtual environment, offering greater expressiveness.
Many works share a similar immersive authoring design but with the aims to lower the technical barriers to program VR / AR applications.
XRSpotlight~\cite{2023FrauXRSpotlight} curates a list of XR interactions from different XR toolkit implementations as natural language rules and help users understand and apply these interactions in a unified way in a 3D scene.
VRIoT\cite{Zhu2023-IoTVR} provides a novel VR learning environment specifically designed to teach students IoT concepts by allowing them to design, program, and explore virtual smart environments with IoT components. 

In parallel to programming interface, a large volume of research works~\cite{zhang2023vrgit,nebeling2021xrstudio,Radu_MacIntyre_2009,Resnick2009-Scratch,leiva2020pronto,Ye2022ProGesAR,Ye2023ProObjAR,Wang2020_CaptureAR,Cho2023RealityReplay,Zhu2022-MechAR,sasaki2013facetons} focus on authoring and prototyping AR contents and interactions.
For example, AR Scratch~\cite{Radu_MacIntyre_2009} is a seminal AR authoring tool for children that enhances the Scratch~\cite{Resnick2009-Scratch} programming platform to help pre-teens create programs that blend real and virtual spaces.
There are also AR prototyping tools helping designers address challenges in prototyping AR interactions, for examples, Faceton~\cite{sasaki2013facetons} proposes a new system for architecture building in an immersive environment, Pronto~\cite{leiva2020pronto} uses a tablet-based interface to enable 3D manipulation and animation, ProGesAR~\cite{Ye2022ProGesAR} supports proximity and gesture-based interactions via a mobile AR system, and ProObjAR~\cite{Ye2023ProObjAR} leverages an AR head-mounted display to facilitate spatial interaction prototyping, with all tools demonstrating enhanced design efficiency and usability in user studies.
Additionally, many works further evaluate the system’s effectiveness and level of immersion during the use of systems using biosignals and EEGs~\cite{cortes2019evaluating,shen2021effects,singh2021impact}.
Recent works also streamline the creation of personalized, context-aware applications. CAPturAR~\cite{Wang2020_CaptureAR} uses an AR head-mounted device to capture and reconstruct daily activities in AR, enabling users to easily create rules and test them instantly, while Teachable Reality~\cite{Cho2023RealityReplay} leverages computer vision method to recognize gestures and interactions with everyday objects, offering a trigger-action interface that simplifies prototyping without programming. 

However, previous works often fall short of fully meeting the needs of advanced users, whose workflow still heavily relies on traditional text-based IDEs such as Visual Studio or Xcode. 
These IDEs provide powerful coding tools and libraries for productive code development and are likely to remain central to program development in the near future. 
\proj{} addresses this gap by offering a new IDE where users can easily program applications around physical objects in a hybrid text and image environment. 
\section{\proj{} System}

\begin{figure*} [ht]
    \centering
    \includegraphics[width=\linewidth]{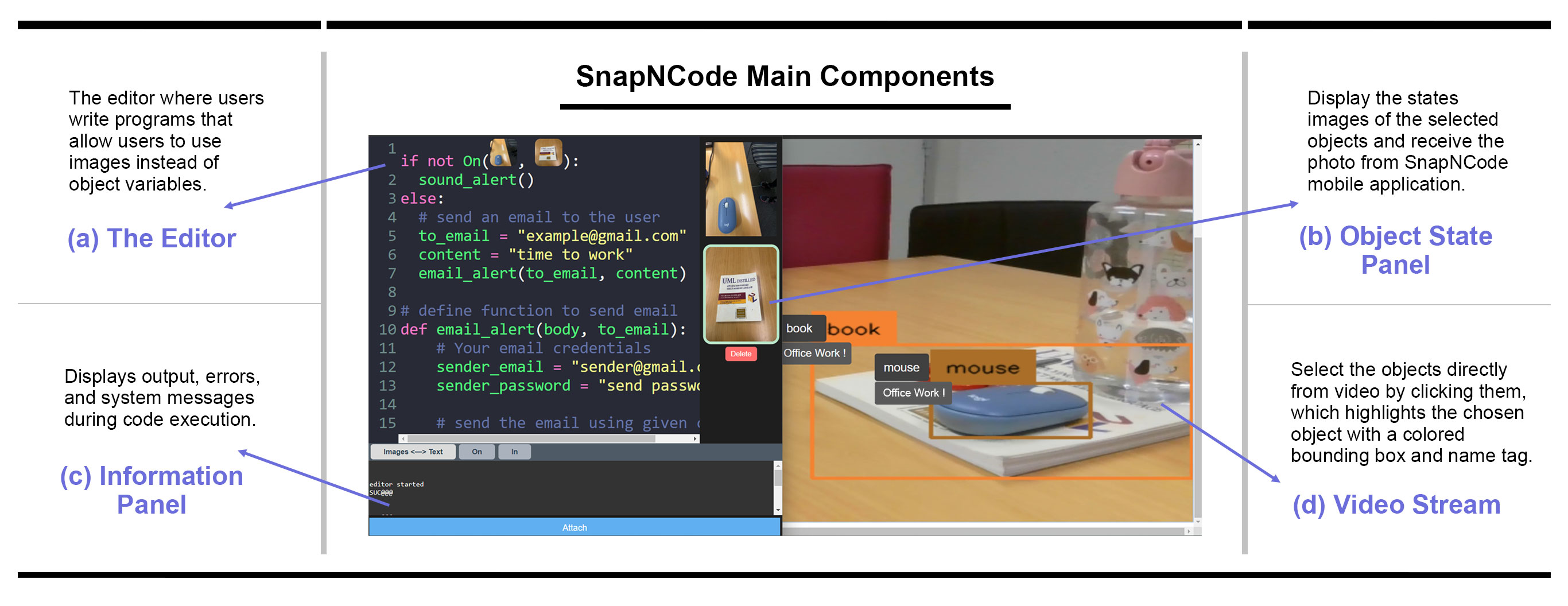}
    \vspace{-6mm}
    \caption{Four main components of \proj{}. }
   \label{Main Interface}
\end{figure*}

\begin{figure*} [ht]
    \centering
    \includegraphics[width=\linewidth]{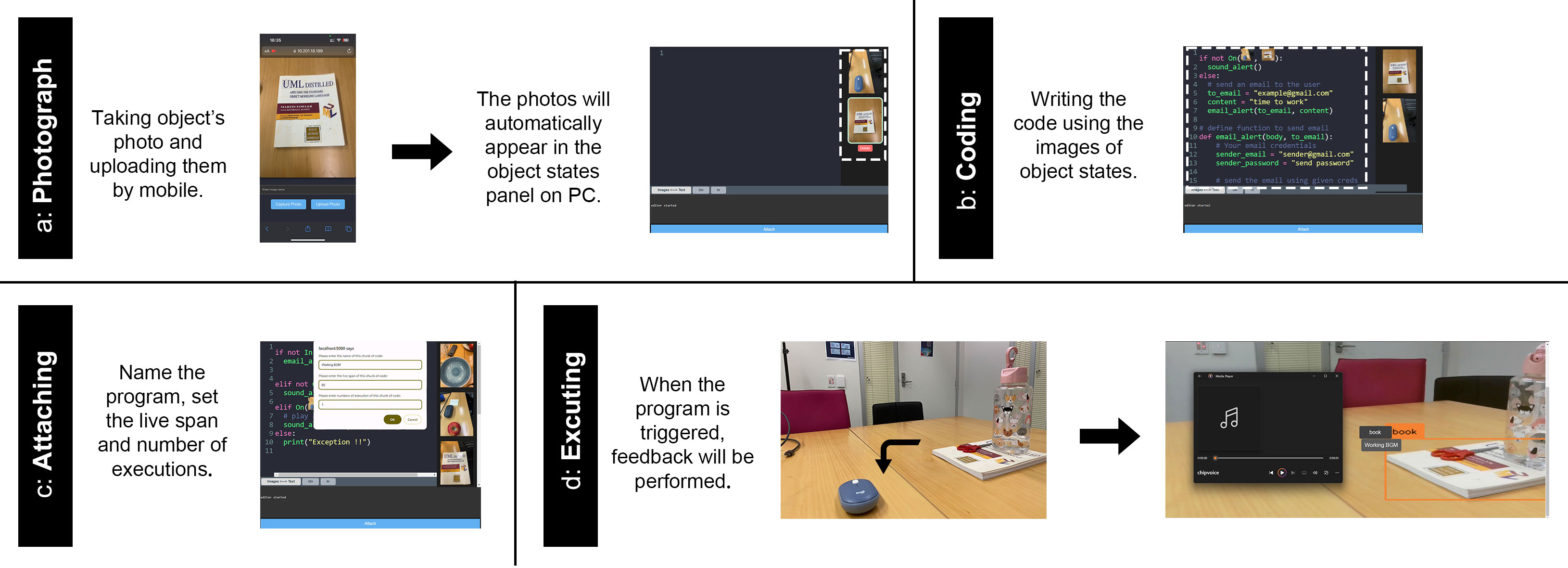}
    \caption{Workflow of \proj{}: Program the different states of the objects: a. Taking the photos and upload by mobile. The photos will automatically appear in the object states panel on PC b. Program Scripting with Python. c. Program set up. Attachment to the Object. d. Triggered Program Execution}
    \label{figure:workflow}
\end{figure*}

\proj{} IDE comprises four main components: (a) the editor, (b) the object state panel, (c) the information panel, and (d) the video stream (Figure \ref{Main Interface}). Here we briefly describe the functionality of each components through a simple scenario: automate the background music, i.e. when the user takes the mouse away from the book, indicating she is starting to work, the music starts. (Figure \ref{figure:workflow}).

Here we briefly describe the functionality of each components through a simple example in the user's daily office work. 

\begin{enumerate}
    \item The user first capture photos of the mouse and book using her phone. If a photo of an object is taken and uploaded successfully, the photo will be displayed immediately in the object status panel (Figure \ref{figure:workflow}a). If the user uses the web camera connected to the PC to identify and click the object in the Video stream, a bounding box will appear around the object.
    \item The user can now directly insert these object states into the code by clicking the mouse and book states in the Object State Panel. Here the user uses a special distance function \textbf{On} in the code to verify if one object is over another. The code in (Figure \ref{figure:workflow}b) specifies that when the mouse is removed from the book, it signals that the user is preparing to start working, thereby triggering the playback of background music
    \item After programming is completed, the user clicks the \textit{attach} button below the text editor. A UI will then appear allowing the user to customize the name of the program, the life span of the program, and the maximum number of times the program is expected to run within that time (Figure \ref{figure:workflow}c). 
    For example, the user inputs 60 and 1, which means that the program runs for 60 minutes. During the period, the program can be triggered up to 1 time.
    \item The code is triggered and executed when the mouse moves away from the book. The code indicator will be automatically removed after its execution (Figure \ref{figure:workflow}d).
\end{enumerate}

\subsection{\proj{} IDE Features}

\proj{} enhances the traditional web-based IDE by introducing compatibility with images, code, and highlighting functions. Additionally, it offers multiple features and UI elements designed to facilitate the creation of spatial computing applications as below:

\noindent\subsubsection*{Snap physical objects into code}

\begin{figure} [h]
    \centering
    \includegraphics[width=\linewidth]{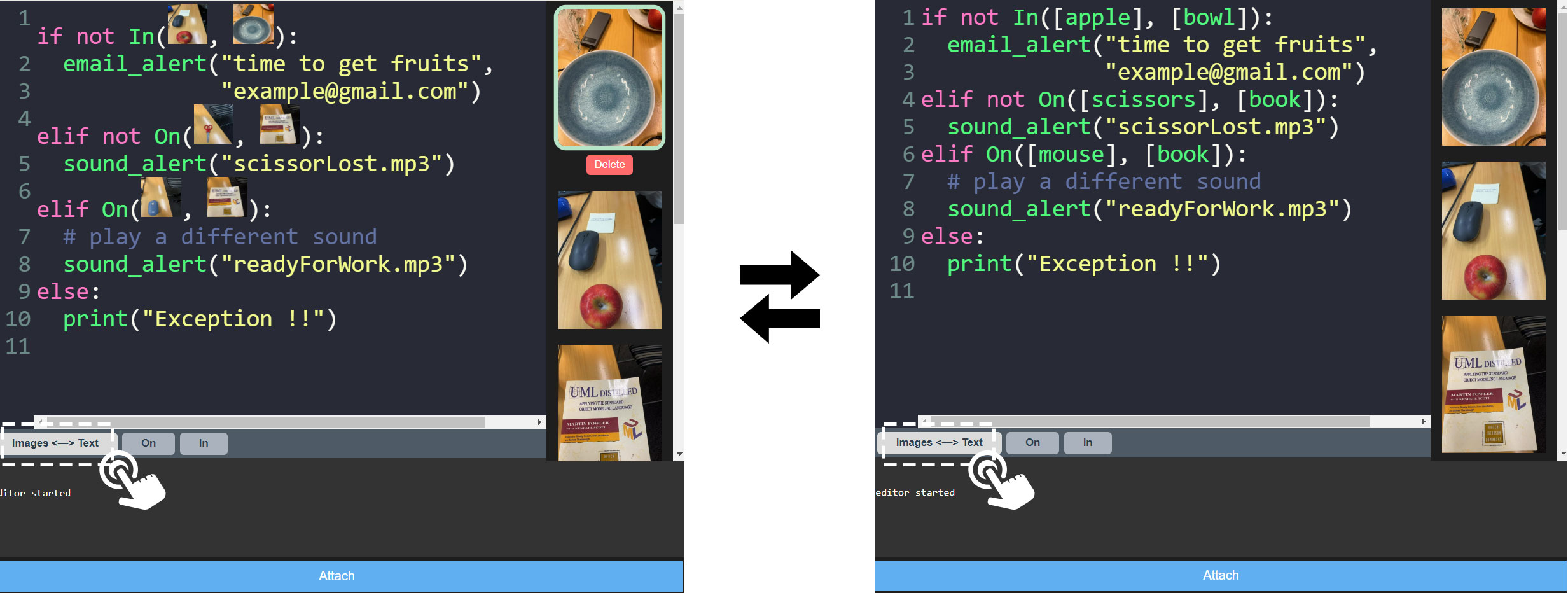}
    \caption{The function of converting embedded images into text descriptors}
   \label{Converting}
\end{figure}

Our IDE enables users to capture physical object states using a web camera or a mobile phone. For instance, a user can take photos of "an open door" or "an empty fruit bowl" using the \proj{} application. These images are then displayed in the Object State Panel of the \proj{} IDE, representing the respective object states (Figure~\ref{Main Interface}).

Users can integrate these states into their code by clicking on the photos in the Object State Panel (Figure~\ref{Converting}). This feature facilitates the creation of event-driven code, such as "play a ringtone when a door is opened" or "add fruits into the shopping list when a fruit bowl is empty".

\begin{figure} [h]
    \centering
    \includegraphics[width=\linewidth]{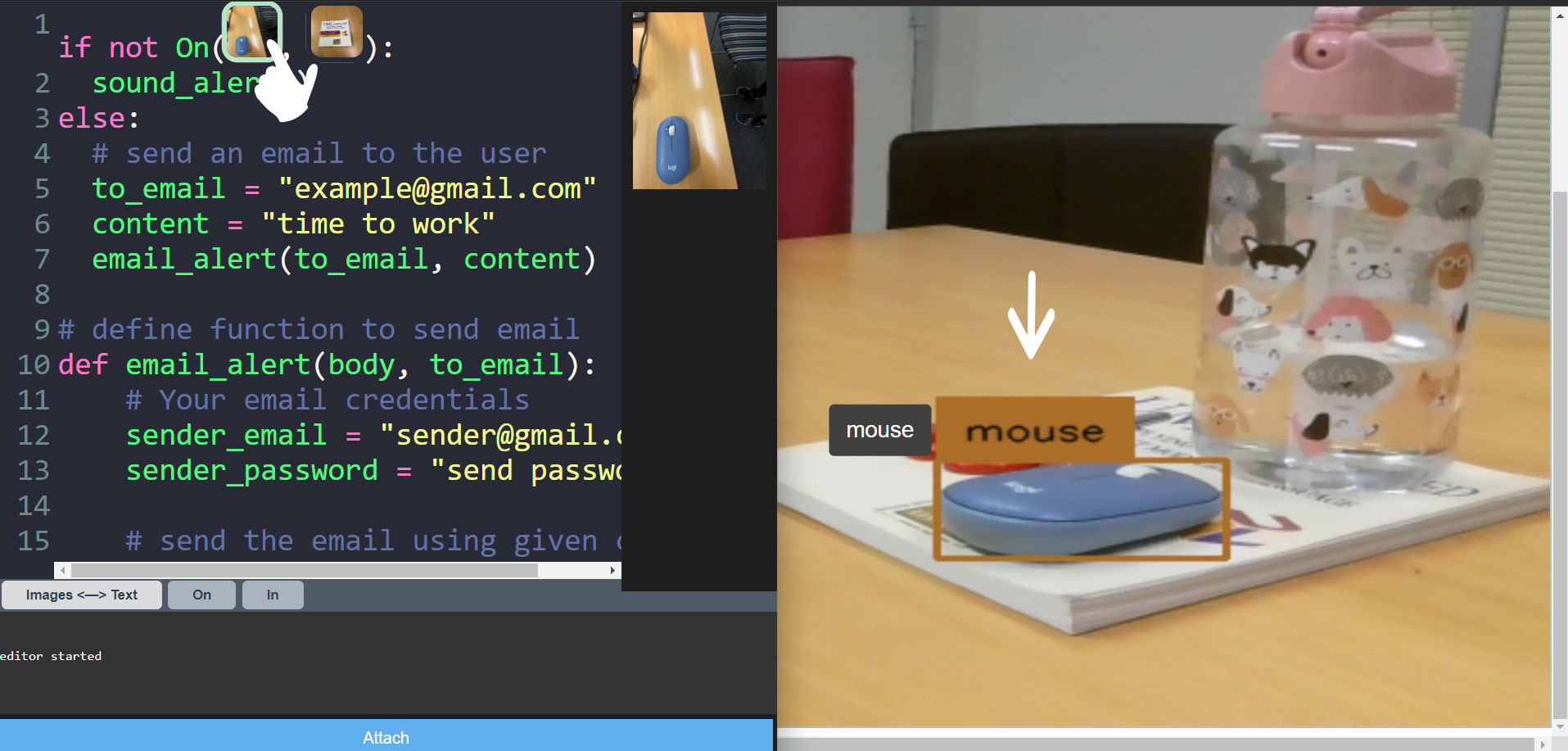}
    \caption{The function of variable highlighting}
   \label{Variable Highlight}
\end{figure}

The visual representation of object states simplifies the process for programmers to mentally link code variables with physical objects. To enhance user understanding of which variable a photo represents, our IDE includes a variable highlighting feature. When the mouse cursor hovers over an inserted object states, the bounding box of its corresponding object in the video stream appears (Figure~\ref{Variable Highlight}).
In addition, in the case where the user would like to investigate the underlying textual representation, \proj{} offers a feature to toggle between images of object states and their text-based descriptions ( Figure~\ref{Converting}).

\noindent\subsubsection*{Attach and Trigger Code}

Upon completing the code, users can \textit{attach} it to a physical object. The attached code snippet will be displayed as a small grey text box adjacent to the physical object (Figure~\ref{Main Interface}d). 
When a video device running the \proj{} application detects the attached code, the code will be triggered and executed (refer to the Implementation section for details).

The code attachment interface (Figure~\ref{figure:workflow}d) allows users to specify the \textit{name}, \textit{lifespan}, and \textit{maximum number of executions} for the attached code. 
These settings determine how long the code remains active before expiring and being automatically removed, as well as its maximum execution count. These features provide users with further control over code execution and help prevent unintended or repetitive triggers.

\noindent\subsubsection*{Additional Utility Python Functions}

The \proj{} offers several customized spatial functions \textit{On()}, \textit{In()}, and \textit{Distance()} relating to the positional relationships between objects. These functions allows coding logic statements such as performing a specific action when two objects exceed a predefined distance.
Please note that our current implementation relies solely on 2D bounding boxes and does not accurately represent precise 3D spatial relationships (see Limitations section).
\section{Implementation}

\begin{figure} [h]
    \centering
    \includegraphics[width=\columnwidth]{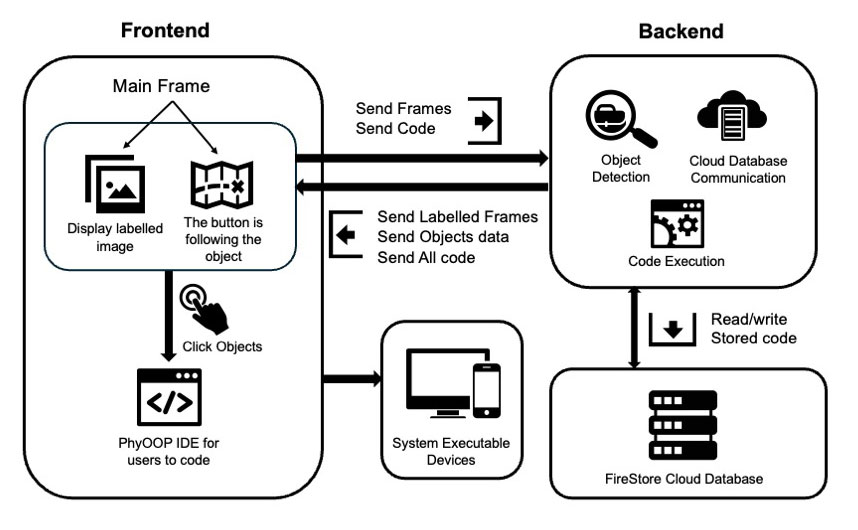}
    \vspace{-5mm}
    \caption{System Design Diagram}
    \label{figure:SystemDesign}
\end{figure}

The \proj{} IDE uses a CodeMirror-based IDE for enhancing coding with image embedding capabilities.
It is a browser-based application designed for compatibility with both desktop and mobile platforms, featuring a dual-component architecture for interaction. \proj{}'s frontend, developed in JavaScript and HTML, offers the \proj{} IDE and sends the captured video frames to the backend for further process. 
The backend uses Python Flask to handle object recognition, video frame display, code execution, and storage/retrieval tasks with the Google Firestore database.

\subsubsection*{Objects Detection}
\label{sec:state_detection}
The \proj{} backend uses a custom-trained YOLOv8 model in conjunction with the official YOLOv8 COCO128 model for instance and object detection.
The former handles the instance tracking and the object categories that are not presented in the COCO128 model (pre-training is required to create a new category for training, please see the Limitation section). 
When the user decides to \textit{snap} an object into the code, a new entry is created in the backend database. This entry includes the object image, its category, and a link to its attached code snippets for future code execution and triggering. 
The tracking information such as the bounding box coordinates and its category is then passed back to the front end for display.


\subsubsection*{Code Triggering}

When a video frame is received by the backend server, it processes the frame using both the custom-trained YOLOv8 model and the YOLOv8 COCO128 model for object detection. If a match is found between the detected object and those stored in the database, the corresponding code is executed on a Python virtula machine in the backend.




\subsubsection*{Spatial Functions}
The \proj{} system introduces spatial computing components, enabling users to address the spatial relationships between objects of interest. Currently, \proj{} offers two distinct spatial functions: the "In" function, which discerns "contained" relationships, and the "Upon" function, designed to identify relationships where one object rests upon another. These functionalities are grounded in a straightforward concept. Leveraging the object detection capabilities of the YOLOv8 engine, the \proj{} system is adept at recognizing real-world objects and extracting pertinent metadata, including the spatial coordinates of each object. By utilizing this location data, the system can accurately compute the "In" and "On" relationships, offering users a deeper insight into the spatial relationship of detected objects.
\section{Usability Study}



The usability study consists of a tutorial, two predefined tasks, one open-ended task, and concluded with an unstructured interview, spanning approximately 90 minutes in total.

\subsection{Participants}
We recruited 12 participants (8 male, 4 female, ages 20-30) from our academic institution. All of them had a programming background, to participate in our tests in groups of 2, for a total of 6 groups. 10 of the participants are adept at utilizing more than one programming language. While 2 of participants are in the nascent stages of learning Python. 
Participants were compensated with a \$20 AUD gift card. 
The study has been approved by the institute's Human Research Ethics Committee. 

\subsection{Introduction and Set-up}
Each participant in the study was provided with a laptop equipped with a webcam and a mobile phone capable of connecting to the \proj{} system. 
The initial introductory session included a simple tutorial, where participants wrote a simple program and attached it onto a computer mouse in the room. 
The program will print "here is a computer mouse" when the mouse appeared in the video stream.


\subsection{Predefined Tasks}
Participants were asked to perform two predefined tasks. For each predefined task, participants will read a task plan and select objects to write programs according to the requirements in the task plan. We prepared for the user: a desk lamp, a book, a mouse, a pair of scissors, and a cup in the experimental area. We chose two simple tasks that will allow the user to explore and use the most of the system Function. The tasks are as follows:
\begin{itemize}
    \item Daily Work Preparation: this scenario simulates the user entering the room and turning on the lamp to start the day. The participant was asked to write a program that automatically opens the day's timetable when the door closes and the lamp is turned on.

    \item Organizing Work Environment: This scenario simulates a user who aims to keep items organized. Participants were asked to write a program that plays a warning sound when stationery is removed from the top of a textbook.
\end{itemize}

\subsection{Open-ended Tasks}
For the open-ended task, participants were asked to collaboratively write a program related to one of two scenarios: office food delivery and retrieval, or joint preparation for a meeting. We pre-trained our model on 10 physical objects, including various stationery items and fruits like apples, bananas, and oranges, which were provided to inspire creativity. Participants received minimal assistance. 

\subsection{Measurements}

All tasks were screen recorded to gather objective measurements such as task completion time, lines of code, and the physical objects involved in the programming. All the participants were invited to fill out a Technology Acceptance Model (TAM)\cite{nielsen1993mathematical} and a System Usability Scale(SUS)\cite{jordan1996usability} questionnaire assessing their experience. 


\subsection{Interview}
Followed the main study, we conducted 15-minute semi-structured interviews with each of our testers on three aspects: 
\begin{itemize}
    \item User interface design and experience: we asked participants about the interface layout and whether it is good looking and practical; and we asked participants about their overall interaction with \proj{}, including experience related to usage fluency, system response speed and feedback effects;
    \item Functionality and Feature Improvements: we inquired about the adequacy of the system's existing features to meet the users' programming needs, any existing functionalities that could be enhanced, and the necessity for introducing new features.
    \item Suggestions for the future: we discussed with the participants the improvement options of the system and envisioned its future application integrated into daily work processes.
\end{itemize}

\section{Result}

\subsection{Overall Experience}

\begin{figure}[h]
    \centering
    \vspace{1pt}
    \resizebox{\columnwidth}{!}{%
\begin{tikzpicture}
\begin{axis}[
    ybar=0.7pt, 
    bar width=18pt, 
    ymin=0, ymax=5,
    ylabel={Average Score},
    xlabel style={at={(axis description cs:0.5,-0.1)},anchor=north},
    ylabel style={at={(axis description cs:0.05,0.5)},anchor=south},
    symbolic x coords={Q1,Q2,Q3,Q4,Q5,Q6,Q7,Q8,Q9,Q10},
    xtick=data,
    nodes near coords,
    x tick label style={rotate=45,anchor=east},
    width=12cm, height=7cm,
    every axis plot/.append style={fill=color3} 
]
\addplot coordinates {
    (Q1,3.6667)
    (Q2,1.9167)
    (Q3,4.0)
    (Q4,3.25)
    (Q5,3.75)
    (Q6,2.75)
    (Q7,3.6667)
    (Q8,2.5)
    (Q9,4.0833)
    (Q10,2.25)
};
\end{axis}
\end{tikzpicture}
} 
    \vspace{-6mm}
    \caption{SUS Average Score}
    \label{fig:SUS Average}
\end{figure}
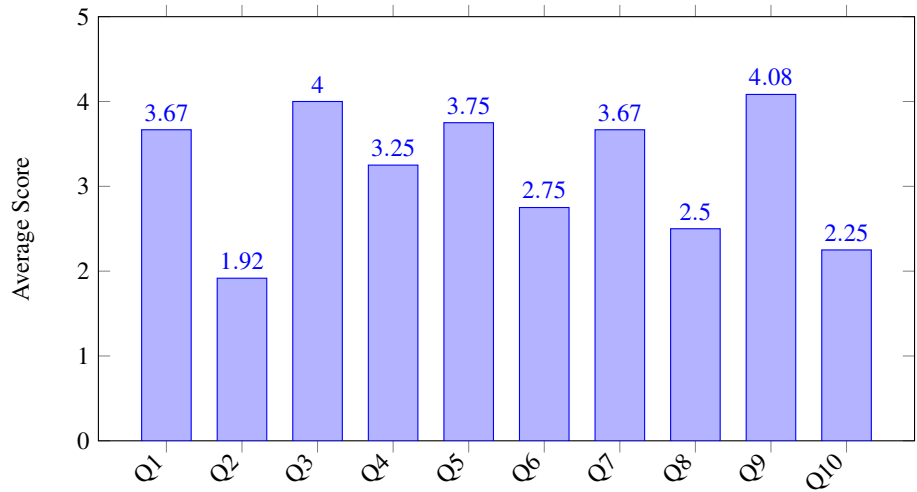

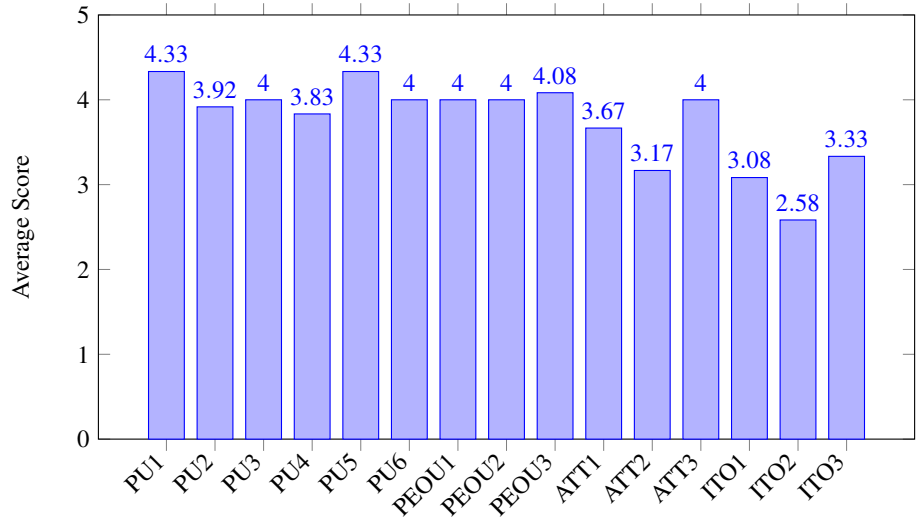
\begin{figure}[h]
    \centering
    \vspace{3pt}
    \resizebox{\columnwidth}{!}{%
\begin{tikzpicture}
\begin{axis}[
    ybar=5pt,
    bar width=13pt,
    ymin=0, ymax=5,
    ylabel={Average Score},
    xlabel style={at={(axis description cs:0.5,0.5)},anchor=north},
    ylabel style={at={(axis description cs:0.05,0.5)},anchor=south},
    symbolic x coords={PU1, PU2, PU3, PU4, PU5, PU6, PEOU1, PEOU2, PEOU3, ATT1, ATT2, ATT3, ITO1, ITO2, ITO3},
    xtick=data,
    nodes near coords,
    x tick label style={rotate=45,anchor=east},
    width=12cm, height=7cm,  
]
\addplot coordinates {
    (PU1,4.3333)
    (PU2,3.9167)
    (PU3,4.0)
    (PU4,3.8333)
    (PU5,4.3333)
    (PU6,4.0)
    (PEOU1,4.0)
    (PEOU2,4.0)
    (PEOU3,4.0833)
    (ATT1,3.6667)
    (ATT2,3.1667)
    (ATT3,4.0)
    (ITO1,3.0833)
    (ITO2,2.5833)
    (ITO3,3.3333)
};
\end{axis}
\end{tikzpicture}
} 
    \vspace{-6mm}
    \caption{TAM Average Score}
    \label{fig:TAM Average}
\end{figure}

Overall, the analysis shows that users are generally optimistic about \proj{} systems. Our System Usability Scale (SUS) assessment provides insight into the perceived usability of a system, as shown in table \ref{fig:SUS Average}. From a sample size of 12 participants, the average score obtained by the system was 66.5 out of 100. The scores indicate that participants are satisfied with the functionality provided by \proj{} and believe that \proj{} can help them effectively and expressively design, demonstrate, and test programs relevant to real-world environments.

As can be seen from the figure\ref{fig:TAM Average}, the score of PU1 is 4.33, indicating that users generally believe that the new programming methods provided by the system help them connect the code with the environment to write useful programs. Indicates that the core functionality of the system is well designed. Next is Perceived Ease of Use (PEOU), with scores showing that participants found the system easy to learn. Attitude towards using the system (ATT), as shown from ATT1 to ATT3, shows a positive trend, with an average score above 3.5, indicating support and satisfaction with the system. There are some fluctuations in the intention to utilize the future system (ITO) divided from ITO1 to ITO3. We conducted interviews with some of the participants above in response to some of the questions raised by them. Try to learn more about user suggestions for the system's user interface, functional experience, and future development.


The predefined tasks measurements indicates that participates were able to efficiently complete the predefined tasks, with the first task averaging 04:32 minutes and the second task, which involved additional steps such as taking photos with the \proj{} mobile app, taking slightly longer at 06:11 minutes. The number of lines of code for each task ranged between 3 to 5.
Errors encountered were mostly spelling mistakes, quickly corrected using the system's information console, which facilitated fast debugging. Some participates also made multiple attempts to experiment with different effects, thus increasing the number of executions. This showcases the system's flexibility and the user's ability to interactively explore and optimize their programming solutions in real-time.

\begin{figure*} [ht]
   \centering
   \includegraphics[width=\textwidth]{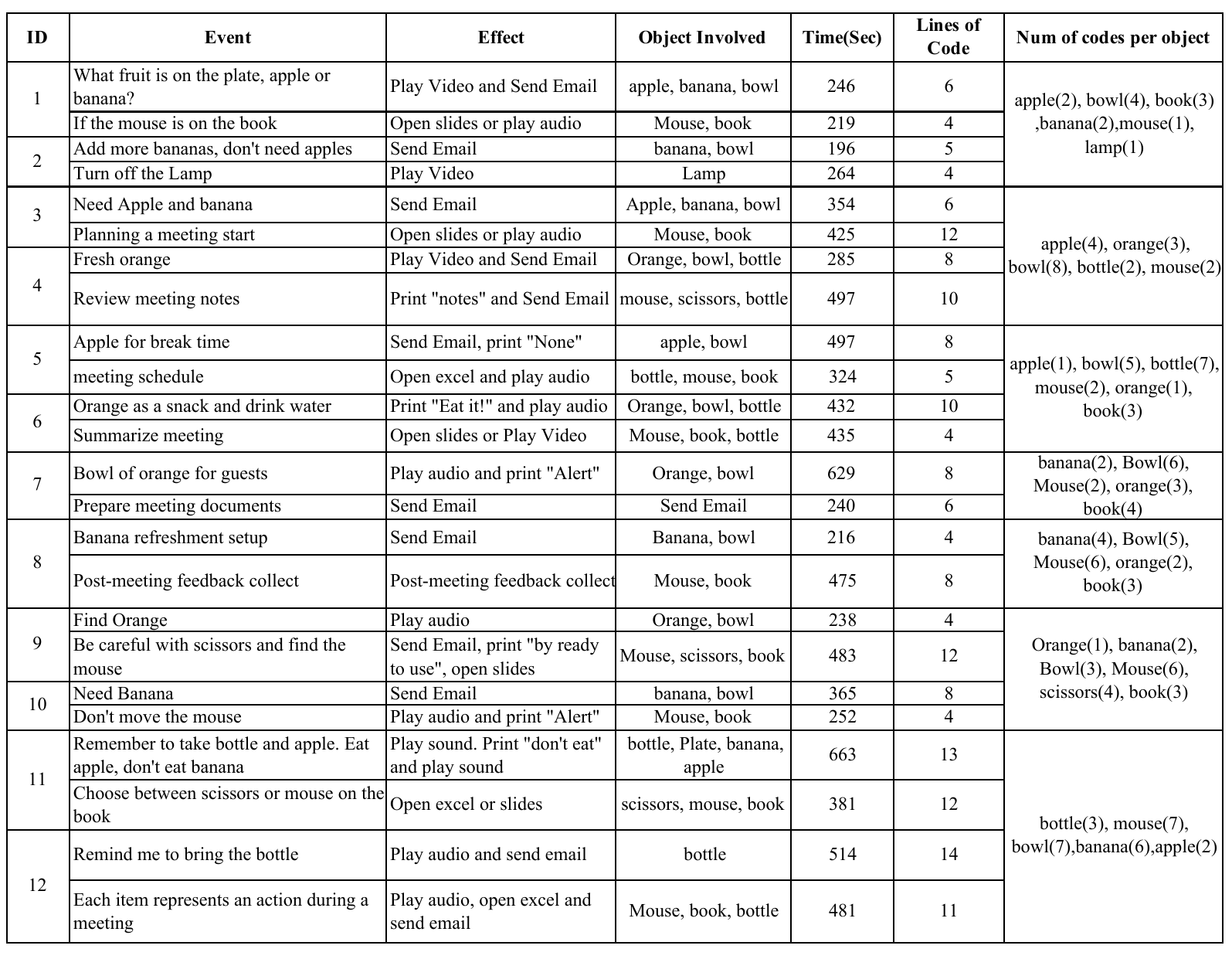}
   \vspace{-2mm}
   \caption{Descriptions of the open-ended prototyping results }
  \label{Table of open-ended taskd result }
  \vspace{-2mm}
\end{figure*}


In the open-ended tasks of the \proj{} project, 12 participants created prototypes that explored various interactive behaviors in real 3D space in pairs. For instance, the system checks whether there is fruit on the plate to trigger events such as playing a video or sending an email, showing in table \ref{Table of open-ended taskd result }. 
From the data, we can find that the task takes between 180 and 700 seconds, and the number of lines of code written ranges from 4 to 16. 
The user time is often related to their programming level and programming complexity. Some testers(P3,P9,P11,P12) are trying to write more complex programs with the system. Participate P9 wrote a complete if elif and else structure, so the program can be used to adopt different scenarios. Participant P12 implemented the SMTP library in Python to enable the sending and receiving of emails directly from the program. However, others(P1,P2,P8,P10) are exploring how to get the job done with minimalist code possible.

These attempts show a high degree of user involvement in the system. It also shows that \proj{} allows participants to transfer many of their programming knowledge in Python to quickly implement their ideas in this new context. Participate P5 and P6 initially programmed \proj{} to trigger a PowerPoint page turn when the mouse was moved away from a book. After observing the interaction, they realized that the command was triggered too frequently, for the mouse was moved frequently by other purposes. To iterate on this design, they added another object in their programs. The PowerPoint slides would now only advance when the mouse was moved away from both the book and the scissors. This addition allowed for more nuanced control, reducing false triggers and aligning the system's responsiveness more closely with intended inputs. 
The prototyping results also highlighted the importance of design and user experience. The fact that participants could quickly iterate their ideas in interaction with real-world spaces demonstrates the utility of \proj{} as a tool for rapid prototyping and user experience testing. 

In our study, open-ended tasks were designed to be completed in pairs, facilitating collaborative programming efforts among participants. Feedback gathered from the eight participants highlighted the significant benefits of working in pairs, such as enhanced brainstorming, efficient exchange of ideas, and the ability to quickly review each other's code. This collaborative framework was commended for creating an environment that encouraged participants to mutually support each other in troubleshooting and refining their code. Furthermore, the remaining four participants agreed that working in pairs was ideal. They expressed concerns that larger groups could complicate the programming process, potentially leading to inadvertent activation of each other's code, thereby increasing the time spent on debugging. 


\subsection{Interview}

\textbf{User interface design and experience}
During the interviews, our participants expressed their views on the user interface and user experience. They like the interactivity of the system, especially in terms of the editor interacting with specific features such as autocomplete suggestions, highlighting status images in correspondence with objects in the video stream, and the information console. 


Participant P6 said, \textit{“What I like most is that the pictures of the state of objects can be uploaded and deleted, because I like to take pictures. Even if there is only one object I will upload many photos.” } In the experiment, through comparison, she deleted many photos which she didn't like of object states. 
Participant P7 thought \textit{“it was great that the code container which stores the code can move with the object. No matter where I put the cup, as long as I can take a photo with my mobile app, I can see the code on it."}

However, the system still needs some improvements. Participant P2 believed that \textit{“it is unnecessary to re-enter all the details every time.” } She believed that there needs to be an interface that can record previous code information.


\textbf{Functionality and Feature Improvements}


Participants generally expressed positive feedback about the system, especially about its interactivity and collaborative potential. They commended the user interface and the real-time object identification capabilities, which significantly enhance the coding experience. However, they also identified areas for enhancement such as the addition of a countdown timer and more explicit feature indicators to assist in discovering and utilizing the functionalities.

Regarding the system functions, two participants P1 and P4 asked for a function that can calculate time. They said: \textit{"Even if I enter the time, I will forget how long is left during the operating system. It would be great if the system can provide a countdown function.”}
Most participants expressed a need for interactive introduction. For example, when the mouse hovers over a function button, a floating window will appear telling the user what this function does. Participant P5 said \textit{“But I wouldn’t have known about the text-to-image feature without being told. Maybe some kind of indicator or hint would help people discover these features.}

In terms of collaboration, participants liked that being able to see other people's code helped them avoid duplicating work\textit{"Being able to see other people's code prevents us from doubling our work."}. Participant P5 believed \textit{"The system could be very useful in everyday life and for professional programmers in larger-scale projects, especially in game design.} He proposed that when multiple programmers work together on a large project, \proj{} can be used as a way to manage code. Programmers can quickly see other people's code and edit it. It can also make code sharing and supervise progress by changing the state of objects.

\textbf{Suggestions for the future}
Participants, including P1, P8 and P11, identified potential applications such as aid for visually impaired individuals or smart home automation.
The incorporation of AI for programming tasks was discussed by Participant P6, considering the use of technologies like GPT for automated code generation.
Participant P2 suggested that the system should appeal not only to programmers but also be simple enough for wider use.
\section{Discussion}

\noindent\textbf{Representation of Physical Objects in IDEs:} The Participants' positive reactions signify the need for more intuitive representations of physical objects within IDEs. Traditional 2D textual representations severely restrict the perceptual and interactive capabilities critical for spatial computing applications. 
\proj{}'s introduction of real-world images as object representations is an innovative first step, but there are more alternatives. 
The proposed approach enables a more natural development of spatial applications by utilizing changes and relationships between physical object states. 
However, achieving nuanced and complex interactions between objects and their environments continues to be challenging and will require more sophisticated tracking and segmentation computer vision algorithms \cite{kirillov2023segment}.

\noindent\textbf{Spatial Context and Coding Environment:} 
An interesting observation from the user study was that many participants prefer to capture the state of physical objects in one environment but then perform the bulk of coding elsewhere, e.g. sit down somewhere else in the room and code. 
This finding suggests that while immersive authoring has its advantages, it might not always be optimal, as participants often favor working in a quieter, distraction-free environment to focus on the more complex and detailed coding tasks.
It seems to suggest that while \proj{}'s integration of live video stream aids in understanding and interacting with physical objects, alternative layouts should be considered to further support different coding context. 
Future IDEs for spatial computing should support seamless transitions from live video stream to recorded and synthetic one as the programmers move between different physical environments.
For example, providing tools that allow easy toggling between different views and states of the physical objects at different time \cite{Cho2023RealityReplay} could keep the spatial context intact without losing the coding efficiency.

\noindent\textbf{Collaboration through Spatial Computing:}
The study highlighted that spatial computing is particularly well-suited for multi-user, in-situ, real-time collaboration \cite{Ye2023ProObjAR,10.1145/3607546.3616804,Wang_Bai_Billinghurst_Zhang_Han_Sun_Wang_Lv_Han_2020}. 
Unlike traditional software development, where most modern IDEs are designed to support asynchronous coding by groups of programmers, spatial computing application would require a more interactive approach. 
\proj{}'s feature that allows code to be attached to physical objects has received many positive feedbaack from participants. 
This feature fosters a shared interactive space where developers can collaboratively modify and enhance object behaviors in real-time. 
This capability not only shifts the programming towards a more spatial orientation but also allows collaborative design and debugging. 
Future IDEs for spatial computing should further enhance these collaborative features by incorporating version control, synchronous editing, and conflict resolution mechanisms to maximize the benefits of spatial computing in a multi-developer environment.
\section{Limitations}
As discussed in Section \ref{sec:state_detection}, \proj{} currently employs a custom-trained Yolov8 model for object state tracking. Unfortunately, this model does not always cover the specific objects or states that users intend to incorporate into their code. If users need to track a particular object instance, re-training the model is necessary. 
In addition, the YOLO model operates within the image space, and our spatial distance function only accounts for the 2D distance, specifically the distance between the centers of bounding boxes. This approach does not capture the 3D spatial relationship between objects.
We anticipate that advancements in computer vision algorithms will eventually overcome these limitations \cite{yang2023track}.

Currently, \proj{} utilizes a purely computer vision-based approach, designed to allow the inclusion of all physical objects within the environment. However, \proj{} has not been tailored specifically for interaction with \textit{smart objects}, e.g., those embedded with sensors, software, and communication capabilities. Integrating both \textit{smart} and \textit{dumb} objects could significantly enhance the coding experience by enriching the interactive environment.


Last but not least, the \proj{} prototype is currently a standalone web-based IDE, utilizing a web interface to enhance cross-device compatibility. Despite our efforts, \proj{} still lacks many of the advanced programming features found in popular IDEs like Visual Studio Code. In the future, we envision developing \proj{} as a plugin for these widely-used IDEs, potentially expanding its functionality and user base.
\section{Conclusion}

This paper presents \proj{}, a novel IDE prototype tailored for spatial computing, which uniquely incorporates physical object states into code as both text and images, bridging the perceptual gap between digital and physical realms. It also allows code to be directly attached to physical objects, enabling collaborative use and context-sensitive activation. 
Positive feedback from our user study demonstrates \proj{}'s potential to streamline spatial application development and integrate smoothly with existing programming workflows.
%
%
%
%

\bibliographystyle{splncs04}
\bibliography{custom}





\end{document}